\documentclass[iop]{emulateapj}
\usepackage{amsmath}
\usepackage{float}
\usepackage{hyperref}
\restylefloat{table}
\usepackage{graphicx}
\usepackage{natbib}
\usepackage{verbatim}

\usepackage{color} 
\usepackage{ulem} 

\begin{document}
\title{Densities and Eccentricities of { 139} {\it{Kepler}} Planets  From transit time variations }
\author{Sam Hadden,Yoram Lithwick}
\affil{Department of Physics \& Astronomy, Northwestern University, Evanston, IL 60208, USA \& Center for Interdisciplinary Exploration and Research in Astrophysics (CIERA)}
\begin{abstract}
	We extract  densities and eccentricities of { 139} sub-Jovian planets by analyzing
	 transit time variations (TTVs) 
	obtained by the {\it Kepler} mission through Quarter 12.
	 We partially circumvent the degeneracies that plague TTV inversion
	 with the help of 
	  an analytical  formula for the TTV.
	  From the observed TTV phases, we find that
	  most of these planets have
	  eccentricities  of order a few percent. 
	  More precisely, the  r.m.s.  eccentricity is { $0.018^{+0.005}_{-0.004}$}, and planets smaller
	  than { $2.5R_\earth$} are around twice as eccentric as those bigger  than { 2.5 $R_\earth$}.
	  We also find a best-fit density-radius  relationship  { $\rho \approx  3
	     $ g/cm$^3\times \left(R/3R_\earth  \right)^{-2.3}$}
	  for the  { 56}
	  planets that likely have small eccentricity and hence small statistical 
	  correction to their masses.   Many
	  planets larger than { $2.5R_\earth$} are less dense than water, implying that their radii
	  are largely set by a massive hydrogen atmosphere.  
\end{abstract}


\section{Introduction}
\label{sec:intro}
	 The {\it Kepler} telescope has 
		 provided an unprecedented
	look into the world of extrasolar planetary systems.
        It has
	 detected the transits of  thousands of  planetary candidates,
	 most of which  are smaller than  Neptune  and orbit with periods 
	$\lesssim 100$ days
	\citep{batalha2013planetary,borucki2011characteristics,borucki2010kepler}.
	 A transiting planet's 
	 orbital period is trivially deduced from the time between transits, and its
	 radius from the depth of transit. 
	  But its other physical properties are much more difficult to obtain, including
	  mass and eccentricity.\footnote{
	  Inclinations can be deduced statistically from the relative numbers
	  of transiting planets in different systems.
	 The inclination dispersion is found to be very small-- a few degrees { \citep[e.g.,][]
	 {lissauer2011architecture,figueira2012comparing,tremaine2012statistics,fang2012architecture,johansen2012can,weissbein2012sterile}.}
	  There may be a second population of highly inclined systems, although that is strongly degenerate with the spacing distribution \citep{lissauer2011architecture}.}  
	 Knowledge of a planet's mass---and hence density---can
	   inform us about its composition; and knowledge of its eccentricity can constrain its dynamical
	   history. Both are key clues towards understanding how the surprising planetary systems discovered by {\it Kepler} formed and evolved.

	  A handful of {\it Kepler} planets  have had their masses measured 
	with radial velocity (RV) follow-up   \citep{koch2010discovery,batalha2011kepler,cochran2011kepler,gautier2012kepler,gilliland2013kepler}.\footnote{  After this paper was  submitted,
	\citet{marcy2014masses}  reported an additional 16 {\it Kepler} planets with
	 RV-measured masses.  
	}  
	But RV measurements are  very difficult to obtain for the bulk of low-mass
	{\it Kepler} candidates.  In
	 multi-planet systems, one can take advantage of  the fact that mutual gravitational perturbations alter
	 the times of transit.
	   If these transit time variations (TTVs)
         can be detected, they can  be used to  characterize  planets
         \citep{agol2005detecting, holman2005use,steffen2013transit,steffen2012transit,fabrycky2012transit,nesvorny2012detection,lithwick2012extracting,xie2012transit}.
	TTVs are exquisitely sensitive to the planets' masses and eccentricities, and can probe
	planets that are too far from their star or too small to yield a detectable
	 RV signal.  But they  depend on  planet properties in a non-trivial way
	 and can suffer from important degeneracies, making it a challenge to invert the 
	 TTV signal to infer the planets' properties.
	
    To overcome this challenge,
    we perform the inversion with the help of
      simple analytical
    formulae for the TTV, derived in \cite{lithwick2012extracting}.
    We focus on near-resonant pairs, which have particularly large TTV signals.
   To a good approximation, the TTV of  
    each planet in such a pair is sinusoidal,
            with period 
       $P'/|j\Delta|$ { (which we call the ``superperiod'')},  
    where 
    $\Delta=P'(j-1)/(Pj)-1$
         is the normalized distance to the nearest $j:j-1$ resonance, and
	$P$ and $P'$ are the average orbital periods of the inner and outer planet.\footnote{
We  follow the convention of using primes to denote quantities corresponding to the outer planet.}
    Thus we write the deviation of the transit times from perfect periodicity 
 as
    the imaginary part of
         \begin{equation}
         V\times e^{  i (2\pi  {j\Delta/P'})t}   \ ,
         \label{eq:v}
         \end{equation}
       for the inner planet, 
         and similarly for the outer with $V\rightarrow V'$.
          The complex amplitudes are set by the planets' masses and eccentricities via
	\begin{eqnarray}
	\label{eq:ttvamp}
	\label{eq:1}
		V &=& \frac{P}{\pi} \frac{\mu'}{j^{2/3}(j-1)^{1/3}\Delta}  \left(-f - \frac{3}{2\Delta}Z_\text{free}^*\right) \nonumber \\
		V' &=& \frac{P'}{\pi} \frac{\mu}{j\Delta}  \left(-g + \frac{3}{2\Delta}Z_\text{free}^*\right) 
	\end{eqnarray}
	where $\mu$ is the planet-to-star mass ratio, 
	$f$ and $g$ are order-unity numbers listed in Table 3 of Lithwick et al. (2012),
	and $Z_\text{free}=fz+gz'$ is a linear combination of the complex free eccentricities ($z = ee^{i\varpi}$) of the two planets.\footnote{
	 {
	 The typical free eccentricities that we deduce in this paper  are much larger than the planets'
	 forced eccentricities, and hence nearly equal to the  total eccentricities.  
	 We shall therefore  call the free eccentricity simply the eccentricity, 
	 with the understanding that it is really the free eccentricity that is deduced from
	 TTV measurements. 
	 }
	}
 
	In this paper, we first extract the amplitudes  ($V$ and $V'$)
	 of { 139}
	  {\it Kepler} planet candidates from the
	    transit times up to Quarter 12 published in
	 \cite{mazeh2013transit}.
	We  then invert Eqs. \eqref{eq:ttvamp} to infer the eccentricities and masses
	(and hence densities) of the planets.  
		Because of degeneracies inherent in 
	Eqs. \eqref{eq:ttvamp}, one must make use of statistical arguments to effect the inversion.
	The approach  we take here is largely similar to that  in \citet{lithwick2012extracting}
	and \citet{wu2012density},  who analyze 22 {\it Kepler}  pairs up to Quarter 6.

\section{Extracting the TTV amplitudes}
\label{sec:sample}
\cite{mazeh2013transit} catalog  the transit times 
 of 1960 KOIs 
with high SNR,  using {\it Kepler} data up to  Q12.
We select planet pairs from their catalog that lie sufficiently close to the first-order resonances 2:1, 3:2, 4:3, or 5:4, such that
 $|\Delta| < 0.06$.
This reduces the sample size to  133 pairs, which is around a  third of all pairs of adjacent planets in their list.
For { each planet in a  pair, we fit the  \cite{mazeh2013transit}  transit times  
 with a sum of two components: one linear in time and the other sinusoidal. 
The period of the latter
  is enforced to be the pair's superperiod ($=P'/|j\Delta|$).
Since the superperiod depends on the individual planet periods, which  
in turn depend on
  the linear components of the fits, 
  we perform a non-linear  Levenberg-Marquardt fit.}
  { 
In systems with three or more planets, 
 the TTV of
		a planet with two perturbers is nearly a linear
		sum of the TTV induced by each perturber alone
		(at the corresponding superperiod). Therefore for such systems, we fit
the TTVs for all expected sinusoidal components simultaneously.\footnote{
		 Three of the triples 
		 require special treatment because  the superperiod
		  of the outer pair is nearly the same as that of the inner pair. 
		 For these ``degenerate triples," we exclude the middle planet's 
		 TTV (as marked by a dash in Table \ref{tab:table1}), because the effect of 
		  its two partners 
		  cannot be disentangled. 
		 In addition, we exclude the following systems from further consideration: KOIs 500 and 730  
		 because of multiple degeneracies;  
		   KOIs 262 and 1858 because their expected superperiods are too long;
		   and  KOI738 because it is strongly perturbed by the 9:7 resonance.
		  \label{footnote:deg}
		 }

Table \ref{tab:table1} lists 95 planet pairs in which one or both partners have ``well-detected'' 
TTV's, which we take 
to mean that their
	TTV amplitudes  are inconsistent with
	 0 at the predicted superperiod with  68\% confidence, based on the covariance matrix
	 generated by the fit.  
	 The table also gives  the values of $V$ and $V'$ (amplitudes and phases) and errors (at 68\% confidence).
	 In total, there are 149 well-detected TTV's in the table, which we refer to 
	 as our primary sample.  
	 After accounting for repeated planets (from triples), our primary sample is comprised
	 of 139 distinct planets.
 }

{ Table \ref{tab:table1}  includes
 21 of the pairs} analyzed previously over a shorter time baseline
 \citep{lithwick2012extracting,wu2012density}.
 \cite{xie2014transit}
   extracts the amplitudes for 15 pairs after determining the transit times
   directly 
from the Kepler light curves.  { Seven}  of his pairs overlap with ones in our list; our amplitudes largely agree with his.

\section{eccentricities and masses from ttv}	
	
We seek to extract the planets'  eccentricities and masses from the
values of $V$ and $V'$
by
inverting Eqs. \eqref{eq:ttvamp}.
 However, there is a strong degeneracy:
 for any pair, one may scale down the masses without affecting the TTV, as long as one correspondingly scales up the eccentricities. 
 Nonetheless,  this degeneracy can be partially lifted
  with a large sample of TTVs.  
   We proceed in two steps. 
   
   \subsection{Eccentricity Distribution}
   \label{sec:ecc}
   
In the first step, we work with    
 the  {\it phases} of $V$ and $V'$ to determine the distribution of eccentricities.  The phases
 depend only on the eccentricities, not the masses (Eqs. \eqref{eq:ttvamp}).  
 For a single planet pair, the  phases cannot be used to infer the eccentricities because they
 depend on the unknown orientation of the planets' Keplerian ellipses ($\varpi$).\footnote{In principal, the two TTV amplitudes and two phases can yield 4 quantities: 
 two planet masses and the real and imaginary components of the combined free eccentricity, $Z_\text{free}$.  However, in most systems 
  the  inner
 and outer planets' TTVs should be nearly perfectly out of phase with one another---whether
    $|Z_{\rm{free}}/\Delta|\gg 1$ or $\ll 1$.  
    Hence there are really only three observed quantities given the typical noise in observed TTVs.}
  But the orientation should be random with respect to the line of sight, i.e.,  the complex phase of $Z_{\rm free}$ should be random. 
  Therefore Eqs. \eqref{eq:ttvamp} imply that
 if all the planets have sufficiently large  $e$ ($e\gg|\Delta|$),
the TTV phases 
will be uniformly distributed between -$180^o$ and $180^o$.  Conversely, if the planets have small $e$ 
 then the phases of the inner planets will all be zero, and those of the outer
 will be 180$^o$.\footnote{We define the
TTV phase as $\phi=\angle(V\times{\rm sgn}\Delta)$ for the inner planet, and the same for the outer
\citep{lithwick2012extracting}.
The extra factor of sgn$\Delta=\pm 1$ allows pairs narrow and wide of resonance to be treated
symmetrically.}

	\begin{figure}
		\begin{center}
		 \includegraphics[trim=4 4 4 4, clip, width=0.5\textwidth]{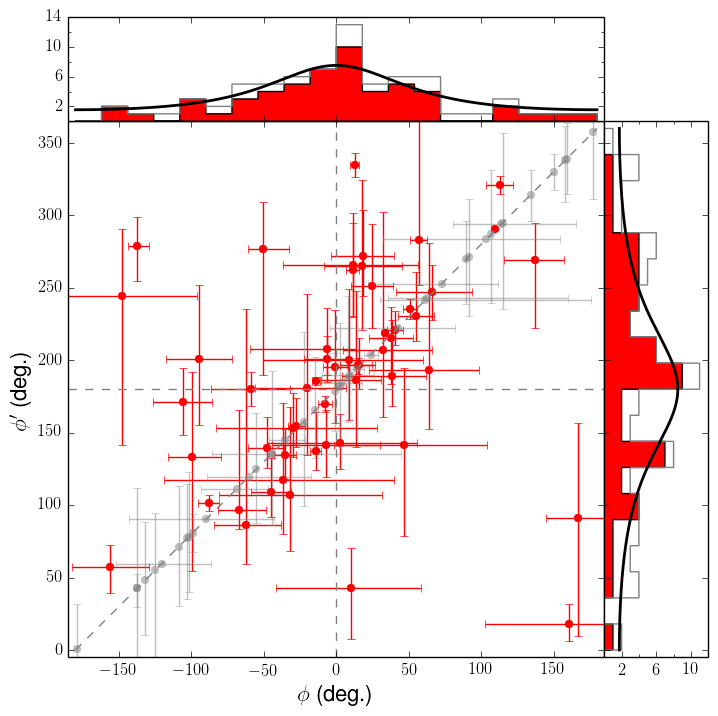}
		 \caption{TTV phases of 
		 planets in our { primary} sample
		(Table~\ref{tab:table1}). 
		 The red points  denote the { 54} pairs in which both planets have well-measured TTVs, and
		 show the inner planet's vs. outer planet's  phase.
		 The grey
		 points denote the {  41} pairs that have a single measured TTV,
		  such that each point (with error bar) corresponds to the detected planet.
		   Error bars denote the 68\% confidence interval.
		 The side panels show  histograms of the red points (shaded histograms) and  grey points
		 (unshaded). 
		 Nearly all the red points  lie along the dashed diagonal
		 line, i.e., the inner and outer planets have anticorrelated phases, as
		  implied by
		  Eqs. \eqref{eq:1}. 
		 Furthermore, the points cluster near, but not always at, the point 
		     $(0^o,180^o)$.  
		     This implies both that $e\lesssim |\Delta|\sim 0.02$ and that it is not true that $e\ll |\Delta|$.
The solid curves in the side panels show the phase distribution produced by our maximum likelihood fit  eccentricity distribution with  
{ $\sigma_e=0.018$}, 
after folding in the observed values of $\Delta$.  
		}
		\label{fig:phihist}
		\end{center}
	\end{figure}

Figure \ref{fig:phihist} displays the phases of our  pairs with well-detected TTV's, along with
histograms of the  inner and outer planets' TTV phases.  The phases unambiguously cluster near (0$^o$,180$^o$)
 indicating that the vast majority of the planets
have $e\lesssim |\Delta|$, i.e., $e\lesssim 0.02$ using the typical value of $|\Delta|$.
But many phases differ markedly from (0$^o$,180$^o$), indicating that many planets cannot have
$e\ll 0.02$.  These inferences are robust, essentially independent of any assumption.

\begin{figure}
\begin{center}
		\includegraphics[width=.45\textwidth]{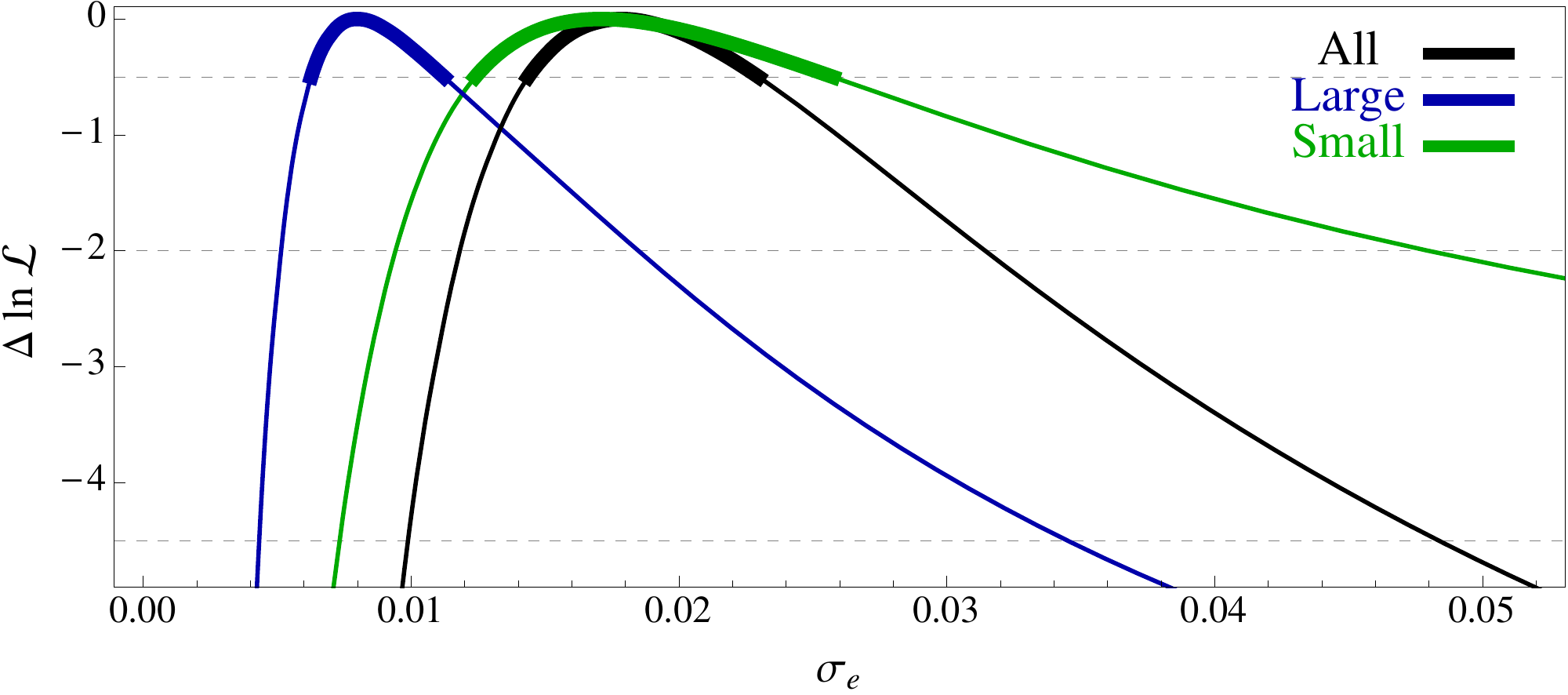}
		\caption{
			{ The dependence of the total likelihood 
			 of  observed TTV phases on  the eccentricity dispersion.
			The vertical axis is the decrease in log-likelihood relative to its maximum.
			Separate curves are shown for all planet pairs (black), large
			 planet pairs with $R~\&\ R' > 2.5R_\earth$ (blue),
			and small planet pairs with $R~\&\ R' < 2.5R_\earth$ (green).
			Dashed horizontal lines indicate decreases in $\ln \mathcal{L}$ of -0.5,-2, and -4.5
			which correspond to nominal $1$, $2$, and $3\sigma$ 
			confidence limits. 
			 }
			}
		 \label{fig:likelihood}
		\end{center}
	\end{figure}

To  be more precise, we assume that the eccentricities
of the planets are drawn from the same underlying distribution, which we take 
to be Rayleigh with scale parameter $\sigma_e $.\footnote{The Rayleigh distribution is equivalent to
a 2D Gaussian $\propto \exp{(-(e_x^2+e_y^2)/2\sigma_e^2)}$ for the real and imaginary parts of the complex eccentricity  $z=e_x+ie_y$.}
   We then determine $\sigma_e$ by  maximizing the total likelihood of the observed phases under the assumed eccentricity distribution.
   The likelihood that a given planet in our sample
   has observed TTV phase $\phi_{\rm obs}$ is the convolution
   \begin{equation}
   \label{eq:LofPhi}
     l(\phi_{\rm obs}|\sigma_e)=\int  P(\phi|\sigma_e)e^{-(\phi-\phi_{\rm obs})^2/(2\sigma_\phi^2)}d\phi \ ,
   \end{equation}
   where $\phi$ is its true TTV phase; the first factor in the integrand is the probability that the phase as determined by Eqs. \eqref{eq:ttvamp}  is $\phi$,
        assuming that the planets' $e$'s are randomly 
   drawn from the Rayleigh distribution; and the second factor is the probability that the noise generates
  the observed phase from $\phi$, where $\sigma_\phi$ is the 68\% confidence error on $\phi_{\rm obs}$
  { (averaging the asymmetric  error bars).}
   Note that $P(\phi|\sigma_e)$ also depends on the (known) value of 
  $\Delta$ but not on the masses. 
  We compute the total likelihood by multiplying together the likelihoods 
  for all  pairs, { where the phase for a pair is taken to be the phase of its inner planet, 
  which in turn is taken to be either $\phi$ or $\phi'-180^o$ depending on which planet 
  has the smaller phase error bar.  }
 Maximizing the total likelihood, we find
  \begin{equation}
	{ \sigma_e = 0.018^{+0.005}_{-0.004} }\ ,
\end{equation}
where the error bars delimit  a decrease in the total likelihood by 
{ $\Delta \ln {\mathcal L
}=-0.5$}
 which would constitute the 68\% confidence interval for a normally distributed  error.
The  solid curves in the side-panels of Figure \ref{fig:phihist} show the phase distribution that results from
the above eccentricity dispersion, using the observed values of $\Delta$. 
 { The dependence of the total likelihood on $\sigma_e$
   is shown in Figure \ref{fig:likelihood}.}

\begin{figure}
\begin{center}
		\includegraphics[trim=0 0 0 0, clip,width=0.45\textwidth]{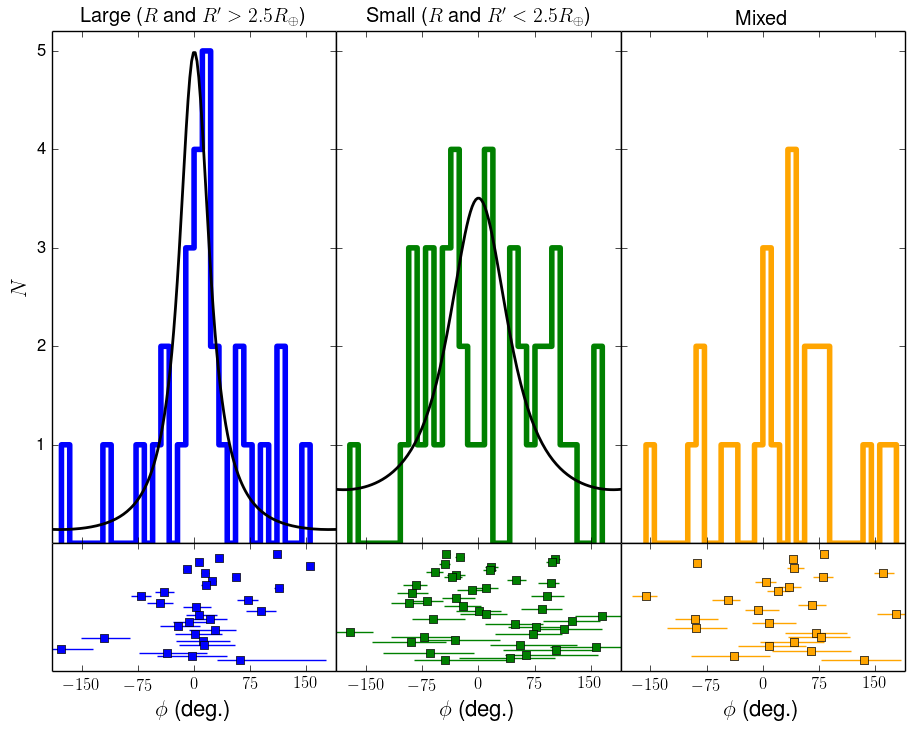}
		\caption{
			Histograms of { inner-planet} TTV phases of the { large} ($R\text{ and }R' > 2.5R_\earth $), 
			 small ($R \text{ and } R' <2.5R_\earth$), and mixed  (one large, one small) planet pairs.
			 { In the lower panels, the individual phases   are shown.}
			The phases of the large planets are more concentrated around $0^o$, indicating they
			have smaller eccentricities.  The phase distributions resulting from 
			the best-fit $\sigma_e$ (Eq. \ref{eq:ss}) are plotted as solid curves { after folding in the observed values of $\Delta$.}
			 }
		 \label{fig:bigVSsmall}
		\end{center}
	\end{figure}

  Figure \ref{fig:bigVSsmall} shows  the distributions of TTV phases for
 large and small planets separately, taking { $2.5R_\earth$} 
 as the dividing line between large and small.\footnote{We detail below how we infer planet radii.}
   { Larger planets tend to have  smaller phases, which are indicative of lower eccentricities.}
 To be quantitative, the phase distribution of pairs of large planets differs from that of  small planets at { 97\% }
 confidence, based on a K-S test  of
the absolute phases.
Fitting for the eccentricity distributions of the two subsamples separately, 
 we find that the large planets 
are around { half }  as 
 eccentric, with eccentricity dispersions { given by}
 \begin{eqnarray}{
  \sigma_{e}=\begin{cases}
                0.017^{+0.009}_{-0.005}
    ,  &{\rm for}\ R\ \&\ R'<2.5R_{\earth} \\
                 0.008^{+0.003}_{-0.002}
              , &{\rm for}\ R\ \&\ R'>2.5R_{\earth} \ ,
      \end{cases} }
      \label{eq:ss}
 \end{eqnarray}
  The best-fit phase distributions corresponding to these $\sigma_e$
are plotted in Figure~\ref{fig:bigVSsmall}. 
 
   A potential concern  is that 
   smaller planets tend to have more uncertain transit times because their transit signal is weaker. Hence 
   their broader distribution of TTV phases
   could be due to measurement error.
   Nonetheless, we feel 
    it likely that smaller planets are truly more eccentric,
    both because
   the error bars in Eq. \eqref{eq:ss} seem sufficiently small to distinguish the two groups, and because another indicator
   of eccentricity---the nominal densities (see below)---also supports
 this conclusion.
  {
  An additional concern is that higher eccentricities boost the TTV signal, making
  it easier
   to detect. So
  while  the $\sigma_e$ we report for small planets with detected TTV's might be correct,
   we are biased against planets with smaller $e$. The same concern does not apply to the large 
   planets because they have $\sigma_e\lesssim |\Delta|$, and hence most cannot have much
   of a boost. 
A more careful investigation of the underlying distributions must await future investigation. 
   }
  
\subsection{Planet Masses and Densities}
\label{sec:dens}
The masses are encoded in the absolute values of $V$ and $V'$.  
	We define the  nominal  mass as  the mass that would be inferred
	 if one assumed (incorrectly, in general)
	 that the { (free)} eccentricity vanishes,
	\begin{align}\label{eq:mnom}
		m_\text{nom} &= M_* \left| \frac{V'\Delta}{P' g} \right| \pi j \nonumber \\
		m'_\text{nom} &= M_* \left| \frac{V\Delta}{P f}\right|  \pi j^{2/3}\left(j-1\right)^{1/3}
	\end{align}
	 (Eqs. \eqref{eq:1}).
	Since eccentricities do not vanish in general,  the true mass $m$ differs
	from the nominal one by an eccentricity-dependent factor: 
	\begin{equation}
	  m={m_{\rm nom}\over|1-{3Z_{\rm free}^*/(2g\Delta)}|}
	  \label{eq:mmnom}
	  \end{equation}
	   with an analogous expression for
	$m'$.
    	If  planets in a pair have  very low eccentricities ($e\ll |\Delta|$), the nominal masses are the true masses, 
	whereas if the eccentricities are much  higher than that ($e\gg|\Delta|$), the
	nominal masses are overestimates of the true masses.
	For intermediate eccentricities ($e\sim|\Delta|$), the nominal and true masses differ
	 by an order unity factor---they can be either smaller or larger depending
	on the phase of 
	 $Z_{\rm free}$.
	 
\begin{figure*}
		\centering
		\includegraphics[width=.80\textwidth]{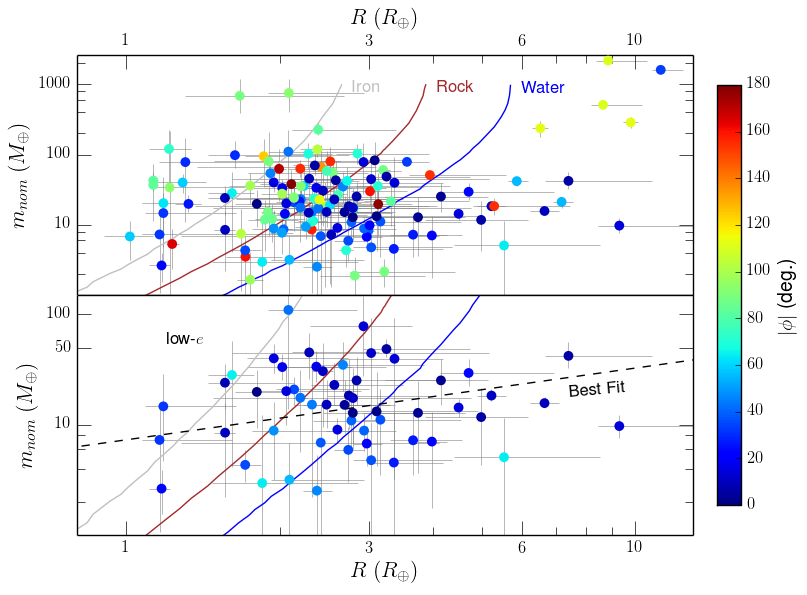}
		\caption{
			{
		Nominal planet mass versus radius.
		The top panel 
		is  }{ the full primary  sample, 
		and the lower panel shows the low-$e$ subsample, for which
		the nominal mass is likely close to the true mass.
		 The best fit relation is given by Eq. \ref{eq:mass}. 
		Colors signify the TTV phase which we take to be
		either $\phi$ or $\phi'-180^o$. depending on which planet in a pair
		has the smaller phase error bar. 
		 Colored curves are theoretical mass-radius relations
		for pure compositions of iron (grey), rock (brown), and water (blue) from \cite{fortney2007planetary}.
		Note the y-axis scales are different in the top and bottom panels.
			}
		}
		\label{fig:mass1}
	\end{figure*}

	{ Table 2 lists the nominal masses of the planets in our primary sample, and the top panel of
	Figure~\ref{fig:mass1} plots these against planet radii.
	For planets with multiple TTV partners, we use the TTV amplitude with
	 the smallest fractional uncertainty to compute its nominal mass.
	 The quoted error  on nominal mass combines the 68\%-confidence error in TTV amplitude with that in stellar
	mass. 
	To determine planet radii and stellar host parameters, we take  star masses and radii  from
	\cite{huber2013revised}, as listed at  the NASA Exoplanet Archive\footnote{http://exoplanetarchive.ipac.caltech.edu/index.html}.
	We  obtain planet radius 
	 by multiplying the star's radius by 
	 the planet-to-star size ratio reported on the Exoplanet Archive; the latter come from
	the literature if available, and otherwise from the Kepler pipeline. 
	There are { 11} host stars among our primary sample for which \citet{mann2013spectro} have published
	revised parameters based on spectroscopic models calibrated to nearby late K and M-dwarfs.
	We use their values of stellar radius and mass along with planet radii where applicable.
	
}

\begin{figure*}
	\begin{center}
		\includegraphics[trim=2 2 2 2, clip,width=.80\textwidth]{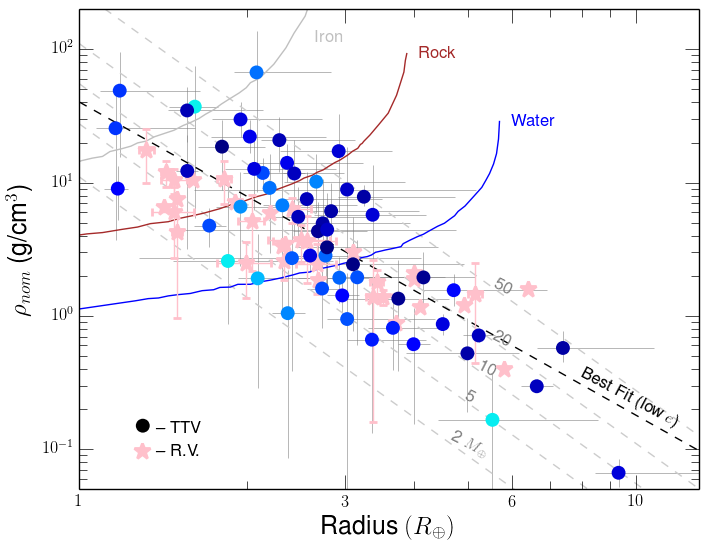}
		\caption{ 
		Nominal planet density versus radius for the { 56} planets in our low-$e$ sample.
		These planets likely have true densities $ \rho \approx \rho_{nom}$.  
		The color scheme for the circles is as described in Figure \ref{fig:mass1}.  The best fit relation  
		is plotted as a black dashed line  (Eq. \ref{eq:dens}). 
		 Pink stars are planets with RV-determined masses, including those listed in Table 3 of \citet{wu2012density}, Kepler planets
		 listed in Table 2 of \citet{marcy2014masses}, as well as
                 Kepler-68c \citep{gilliland2013kepler}, and HD97658b \citep{dragomir2013warm}.
		}
		\label{fig:dens}
	\end{center}
\end{figure*}

\begin{figure}
		\begin{center}
		\includegraphics[width=0.5\textwidth]{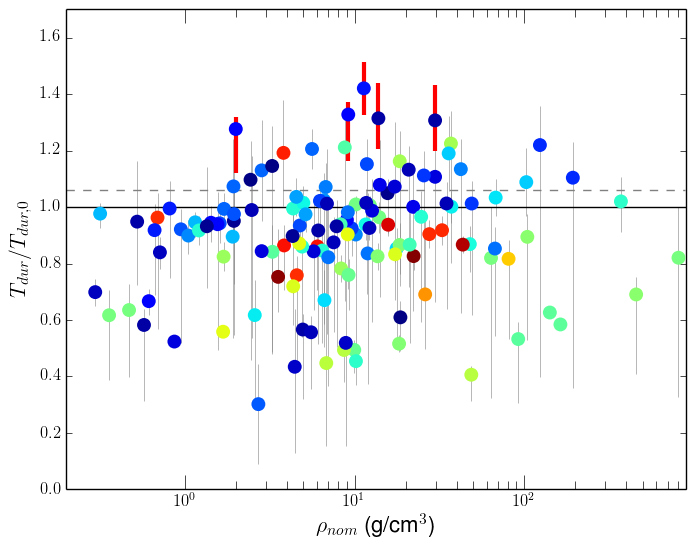}
		\caption{
		 Selecting high-$e$ planets based on  transit duration. The x-axis shows the planets' nominal density, 
		and the y-axis shows the ratio of their
		observed transit durations to what the duration would be if the planet were on a circular orbit with zero impact parameter.  When that ratio exceeds unity,  
		the eccentricity is bounded from below by
		$e>T_\text{dur}/T_\text{dur,0}-1$ \citep[Eq. 1 of][ to linear order in $e$]{ford2008characterizing}.
		The dashed line corresponds to a minimum eccentricity of { $0.06$}, which is greater than 99.7\% 
		of eccentricities belonging to our best-fit 	eccentricity distribution.
		The color/symbol scheme is as described in Fig. \ref{fig:mass1}.
		 Error bars are  based on uncertainty in stellar radius, and are suppressed in the x-direction.
	        { Five planets are selected as high-$e$ based on transit duration alone 
	       and are emphasized with bold red error bars. } 
		 In addition, many of the planets selected as high-$e$ based on TTV phase or nominal density also have large
		 transit durations, corroborating their membership in the high-$e$ sample. 
		}
		 \label{fig:duration}
		\end{center}
	\end{figure}

	In order to focus on pairs that  likely  have small eccentricities, and thus
	 true masses close to their nominal masses,
	we split our sample   into  ``low-$e$'' and ``high-$e$'' planets.
A pair is categorized as  high-$e$  if it satisfies any of the following criteria:
\begin{enumerate}
\item    The TTV phase differs, at 68\% confidence, by more than 30$^o$ from its zero-$e$ value, i.e., from
0$^o$ for the inner planet or 180$^o$ for the outer, using whichever planet has the smaller phase error.

\item 
{ The nominal density exceeds that of pure iron composition at 68\% confidence.  
Since true densities are likely not so high, these planets presumably have large $e$ and hence
 a large correction to their nominal masses.
This  conjecture is supported by the fact that
many   planets nominally denser than iron possess large TTV phase (Fig. \ref{fig:mass1}).}

\item 
 One or both planets have transit durations that imply $e>|\Delta|$ at 68\% confidence (Fig.~\ref{fig:duration}).
 { This criterion could be triggered by an underestimated radius for the planet's
 host star, rather than the planet's high $e$.  But for such planets, the nominal density would also exceed
 the true density.}
 \end{enumerate}
We find { 83} planets  fall into the high-$e$ sample, 
{ and { 56}  in the low-$e$ sample. 
	 The nominal masses for }
	{ the latter are shown  in the lower panel of  Fig. \ref{fig:mass1}, and
	the nominal densities in Fig. \ref{fig:dens}.}

 { We determine a mass-radius relationship for the low-$e$ planets\footnote{
   We exclude Kepler-51c (KOI620.03) from our 
mass-radius fits, as it is significantly larger  than the other planets ($9.3R_\earth $).}
}{ by performing a linear least-squares fit in log-log space, yielding}
\begin{equation}\label{eq:mass}
	{	m_{\rm nom} \approx 14.9^{+3.4}_{-2.8}~M_\earth~\times {\left( \frac{R}{ 3 R_\earth} \right)}^{0.65\pm0.14}}
\end{equation}	
{ The nominal masses likely exceed the true masses by a modest amount for the low-e
sample. Applying the statistical correction procedure described in  \cite{wu2012density}, 
we estimate that the correction factor to the best fit is { $\lesssim 10\%$}, and hence
we ignore it.  
Converted to densities, the above fit implies}
\begin{equation}\label{eq:dens}
{ \rho_{\rm nom}\approx  3  \mathrm{~g/cm}^3\times \left(R/3R_\earth  \right)^{-2.3} \ .}
\end{equation}
 { This  agrees reasonably well with 
 what was   found in \cite{wu2012density} based on a smaller sample of TTV's. It is also 
 in adequate agreement with
 results from planets with RV-determined masses (Fig. \ref{fig:dens}), which were not included in the fit.
 \citet{weiss2014mass} report}
{
 $m \approx 7~M_\earth  \times (R/ 3 R_\earth)^{0.93} $, for planets in the size range
 $ 1.5R_\earth < R < 4R_\earth$, based primarily on RV measurements.
}	
 
\begin{figure}
		\begin{center}
		\includegraphics[width=0.45\textwidth]{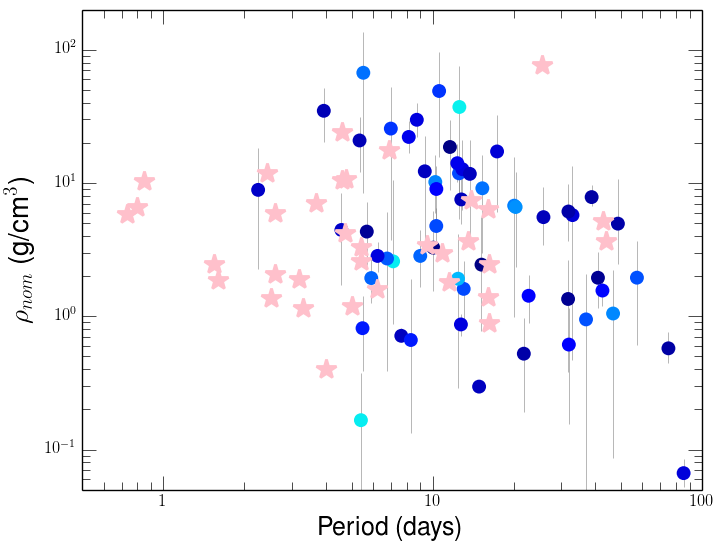}
		\caption{  
			Nominal planet density { of the low-$e$ subsample}
			versus orbital period.  The color  and symbol scheme is the same as described in Fig.~\ref{fig:mass1}.
			While planet densities are quite scattered,  there is a hint of
			closer-in planets  being denser.
			The RV planets probe significantly shorter periods than the majority of  TTV
			planets.  }
		 \label{fig:rhoVSp}
		\end{center}
	\end{figure}

 Large planets ($>2.5R_\earth$) appear to be distinct from small ones ($<2.5R_\earth$)  in a number of ways.
     First,
     larger planets tend to be less dense.
       Many large ones 
       are less dense than water, indicating that
their radii must be set to a large extent by a gaseous hydrogen atmosphere.
 By contrast,  nearly all small ones
are denser than water, and some are even denser than rock. 
Second, small ones tend to be closer to their star: the
 median orbital period
 for small planets in our  { primary sample}
  { is  10.1~d }, whereas
that of the large ones is { 15.3~d}.  
Such a correlation for Kepler candidates has been noted in \cite{wu2012density} and \cite{owen2013kepler}. As argued there,  
all planets might have started out with gaseous envelopes, and then the ones closer
to their star were photoevaporated.
However, this correlation could also be owed, at least partially, to biases in transit detectability \citep{gaidos2013objects}.
We also find here modest evidence that closer-in planets are denser
(Fig. \ref{fig:rhoVSp}).

 And third, as shown above,  small ones  are  around 
 twice as eccentric (Eq. \ref{eq:ss}; Fig. \ref{fig:bigVSsmall}).
 This is corroborated by the fact that planets  
 in Fig.  \ref{fig:mass1} with nominal densities higher than iron
  have  $R<2.5R_\earth$. 

It would be of interest to know whether {\it Kepler}
planets are mostly water. If so, it would suggest that the planets
formed beyond  the ice-line and then migrated inwards; if not, it  would argue for {\it in situ}
formation.    Unfortunately, the water content is difficult to deduce, 
since the density is very sensitive to the presence of hydrogen; e.g., a
 $5M_\earth$ rocky planet with 
a hydrogen atmosphere that is only $\sim 1\%$ of its mass would have the same density as a pure water
planet
 \citep{adams2008ocean,wu2012density}.
But we conjecture that the majority of planets are nearly water-free, based on the fact that planets of a given mass span such a wide range of densities---from less dense than water to denser than rock, 
  without much evidence for a pileup of planets near the
water-density curve.  Furthermore, the fact that small planets tend to be closer to their star  suggests
that those might have lost their gaseous envelope by photoevaporation, exposing the rocky core underneath \citep{lopez2012thermal, wu2012density, owen2013kepler}.

\section{Summary}

We have extracted the eccentricities  and masses---and hence densities---of   a large sample of  
 {\it Kepler} planets, starting from the  \cite{mazeh2013transit}
  catalog of transit times through Quarter 12.
  We found { 139} planets suitable for our analysis.
In order to invert the observed TTVs, we used an analytical formula for the TTVs of
 near-resonant pairs.  Inverting the analytical formula is not only  much faster than the more typically used N-body inversion  \citep{holman2010kepler,lissauer2011closely,cochran2011kepler,lissauer2013all}, 
but it also allows one to break the strong degeneracy between mass and eccentricity in 
a statistical way.
We largely followed the approach of \cite{lithwick2012extracting,wu2012density}, but applied it to
a significantly larger sample and for nearly twice the duration.
Our main results are as follows.
\begin{itemize}
\item From the TTV phases, the planets' eccentricities are of order a few percent, with r.m.s. { $\sigma_e=0.018^{+0.005}_{-0.004}$ },
 when fit with a Rayleigh distribution.  Small planets ($<2.5~R_\earth$) are { around twice}  as eccentric as the larger ones.
\item The nominal masses---and hence densities---were extracted from the TTV amplitudes.
{ The nominal densities of the low-$e$ planets, which are expected to be 
 comparable to the true planet densities, 
are  displayed in Fig. \ref{fig:dens} (see also  Table \ref{tab:table2}).
The mass-radius relation we infer from TTV's largely agrees with that found
from RV.}
Large planets ($> 2.5~R_\earth$) appear distinct from small ones, in
 that many of them are so underdense that they must be covered in gas,
  whereas most small ones are as dense as water or rock.  
In addition, the large ones are typically less eccentric and  further from their host star.
\end{itemize}

The untimely malfunction of the Kepler satellite
makes it unlikely that
the number of planets with TTV-measured
eccentricities and densities 
will increase substantially
in the near future. 
But there are a
number of prospects for improving upon the results of this paper. 
First, {\it Kepler} has released 
lightcurves up to Quarter 16, and the present study may
be extended to  incorporate them.
{ Second, the radii of stars in the Kepler catalog 
are uncertain. This
directly translates into uncertainty in the 
planet radii---and hence  densities.
For a crude estimate of the error, we compare the radii of the 38 stars in our primary sample
 that have been followed up (e.g., with spectroscopy or asteroseismology) with their pre-follow-up values, and find
that the median error is 12\%; 31 of the stars have errors $ < 25\%$, but three have errors larger than 50\%.
 In the future, more accurate measurements of stellar radii will yield more accurate planet densities.}

Third, the TTV signals sometimes have components in addition to the dominant one of
Eq. \eqref{eq:v}, especially when a planet pair lies near a  higher-order resonance.  If these are detected and decoded, they can be used to help break
the mass-eccentricity degeneracy. And fourth, RV measurements  will continue to expand
our knowledge of sub-Jovian planets.

\section*{acknowledgments}
	We thank { Will Farr, Eric Gaidos, Jason Steffen, and Yanqin Wu }  for helpful discussions.
	We are also grateful to the {\it Kepler} team for acquiring and publicly releasing such  spectacular results.
	YL acknowledges support by 
	NSF Grant AST-1109776 and NASA grant NNX14AD21G.
\bibliographystyle{apj}
\bibliography{KeplerTransitTTV}
\clearpage
\begin{deluxetable*}{ccccc ccccc cc}
 \tabletypesize{\scriptsize}
\tablecaption{TTV Amplitudes}
\tablewidth{0pt}
\tablehead{
	\colhead{KOI}  &\colhead{$P$} &\colhead{$P'$} & \colhead{$j$}& \colhead{$\Delta$}&\colhead{$|V|$}	&\colhead{$|V'|$ }&\colhead{$\phi$ }& \colhead{$\phi'$ }\\ 
\colhead{in/out} & \colhead{(d)}& \colhead{(d)}&& &\colhead{min.}&\colhead{min.}  &\colhead{deg.} &\colhead{deg.}
	}
\startdata
82.02/01&	10.31&	16.15&	3&	0.044&	--&	$ 1.4^{+0.5}_{-0.5} $&	--& $ 159^{+25}_{-25} $ \\ 
82.04/02&	7.07&	10.31&	3&	-0.028&	$ 16.0^{+5.8}_{-5.8} $&	--&	$ 10^{+19}_{-17} $& -- \\ 
85.01/03&	5.86&	8.13&	4&	0.041&	$ 0.2^{+3.6}_{-0.2} $&	$ 4.1^{+2.8}_{-2.8} $&	$ -33^{+180}_{-180} $& $ -102^{+38}_{-45} $ \\ 
111.01/02&	11.43&	23.67&	2&	0.036&	$ 1.1^{+2.3}_{-1.1} $&	$ 3.3^{+2.1}_{-2.1} $&	$ 5^{+180}_{-180} $& $ 2^{+42}_{-43} $ \\ 
115.01/02&	5.41&	7.13&	4&	-0.013&	$ 1.3^{+3.0}_{-0.8} $&	$ 3.1^{+2.5}_{-2.5} $&	$ -148^{+103}_{-46} $& $ -116^{+54}_{-52} $ \\ 
137.01/02&	7.64&	14.86&	2&	-0.028&	$ 5.3^{+0.4}_{-0.4} $&	$ 4.1^{+0.3}_{-0.3} $&	$ -8^{+5}_{-5} $& $ 170^{+5}_{-5} $ \\ 
148.01/02&	4.78&	9.67&	2&	0.012&	$ 4.0^{+1.2}_{-1.2} $&	$ 3.1^{+0.8}_{-0.8} $&	$ -6^{+16}_{-16} $& $ -159^{+16}_{-16} $ \\ 
152.02/01&	27.40&	52.09&	2&	-0.050&	$ 1.0^{+4.0}_{-1.0} $&	$ 5.7^{+1.3}_{-1.3} $&	$ 17^{+180}_{-180} $& $ -99^{+14}_{-13} $ \\ 
152.03/02&	13.48&	27.40&	2&	0.016&	$ 8.2^{+2.8}_{-2.8} $&	$ 21.7^{+3.9}_{-3.9} $&	$ -36^{+17}_{-18} $& $ 135^{+8}_{-8} $ \\ 
156.01/03&	8.04&	11.78&	3&	-0.024&	$ 1.3^{+1.4}_{-1.3} $&	$ 3.3^{+0.7}_{-0.7} $&	$ -5^{+180}_{-180} $& $ 150^{+12}_{-12} $ \\ 
...&...&       ...&...&...       &...&...     &...&...   
\enddata
	\tablecomments{
	{ Table \ref{tab:table1} is published in its entirety in the electronic edition of {\apj}. A portion is shown here for guidance. 
	Each line gives the TTV amplitudes for the inner and outer member of a pair of interacting
	planets, obtained by fitting the transit times catalogued in  \cite{mazeh2013transit}. 
	The TTV amplitudes (complex numbers $V$ and $V'$) are denoted here by
	their magnitudes
	($|V|$ and $|V'|$) and phases ($\phi$ and $\phi'$), with
	error bars  at 68\% confidence.
	Our primary sample consists  of  planets whose TTV magnitudes are inconsistent
	with zero within the error bar. 
	 In other columns, 
    $P$ and $P'$ are  orbital periods,  j is the nearest  j:j-1 resonance, and
	$\Delta$ is the normalized distance to  resonance.  Dashes appear for middle planets in
	 degenerate triples (see footnote \ref{footnote:deg}).
	 }
	}
	\label{tab:table1}

\end{deluxetable*}

\begin{deluxetable*}{cccccc}
 \tabletypesize{\scriptsize}
\tablecaption{Planet Nominal Masses}
\tablewidth{0pt}
\tablehead{
	\colhead{Planet} & \colhead{$m_\text{nom}\tablenotemark{a}$} & \colhead{$R_p$}    	   &\colhead{$e$-flag\tablenotemark{b}} & \colhead{$M_*$} &\colhead{$R_*$} \\ 
			      &	 \colhead{($M_\earth$)}       & \colhead{($R_\earth$)} & \colhead{(h/l)}& \colhead{($M_\sun$)} &\colhead{($R_\sun$)} 
	}
\startdata
82.02&	$ 2.6^{+1.3}_{-1.1} $&	$ 1.2^{+0.1}_{-0.1} $&	l&	0.73& 0.755 \\ 
85.01&	$ 26.6^{+20.4}_{-18.5} $&	$ 2.6^{+0.04}_{-0.04} $&	h&	1.27& 1.424 \\ 
111.01&	$ 19.6^{+14.5}_{-12.4} $&	$ 3.1^{+0.5}_{-0.7} $&	h&	0.81& 1.361 \\ 
115.01&	$ 5.1^{+6.3}_{-4.1} $&	$ 5.5^{+3.0}_{-1.1} $&	l&	1.09& 1.332 \\ 
115.02&	$ 3.0^{+9.2}_{-2.0} $&	$ 1.9^{+1.0}_{-0.4} $&	l&	1.09& 1.332 \\ 
137.01&	$ 18.4^{+2.7}_{-1.9} $&	$ 5.2^{+0.4}_{-0.3} $&	l&	0.88& 1.055 \\ 
137.02&	$ 15.7^{+2.0}_{-1.4} $&	$ 6.6^{+0.5}_{-0.4} $&	l&	0.88& 1.055 \\ 
148.01&	$ 14.3^{+4.3}_{-4.5} $&	$ 2.0^{+0.2}_{-0.1} $&	h&	0.96& 0.852 \\ 
148.02&	$ 9.8^{+3.3}_{-3.4} $&	$ 3.0^{+0.4}_{-0.1} $&	h&	0.96& 0.852 \\ 
152.02&	$ 13.7^{+6.5}_{-4.5} $&	$ 2.0^{+0.9}_{-0.2} $&	h&	1.08& 1.037 \\ 
...	  &				...	&		...			&	...&	...	&	...
\enddata
	\tablecomments{
	{
	Table \ref{tab:table2} is published in its entirety in the electronic edition of {\apj}. A portion is shown here for guidance.
	\tablenotetext{a}{ Planet nominal mass, given by Eq. \eqref{eq:mnom}}
	\tablenotetext{b}{Flag indicating whether the planet is classified as high-$e$ (``h") or low-$e$ (``l") according to criteria listed in Section \ref{sec:dens}}
	}
	}
\label{tab:table2}

\end{deluxetable*}

\end{document}